\documentclass[a4paper,11pt]{article}
\usepackage{pos,amsmath,url,cancel,lineno}

\newcommand{\vect}[1]{\mathbf{#1}}
\newcommand{\ttbar}{\ensuremath{t\overline{t}}}

\newcommand{\Powheg}{{\textsc{Powheg}}} 
\newcommand*{\PYTHIA}{{\textsc{pythia8}}}
\newcommand{\Hathor}{{\textsc{HATHOR}}}

\title{Measurements of top quark properties in CMS: \ttbar~spin density matrix, quantum entanglement and quantum magic}
\ShortTitle{Top quark properties in CMS}

\author*[a,b]{Efe Yazgan}

\affiliation[a]{on behalf of the CMS Collaboration}

\affiliation[b]{National Taiwan University,\\
  Department of Physics, Laboratory of High Energy Physics, 10617 Taipei, Taiwan}
  
\emailAdd{efe.yazgan@cern.ch}

\abstract{
Polarization and spin correlation measurements of top quark-antiquark (\ttbar) pairs provide tests of the standard model, but also new ways to test quantum mechanics with unstable particles at highest energies ever produced in a laboratory. 
Recent \ttbar~spin correlation measurements and the tests they enable, made with the CMS detector at the CERN LHC Run 2, are presented. The measurements summarized include the full spin density matrix measurement of top quark pairs using events with a single lepton and jets in the final state. 
Spin correlation measurements in specific phase space regions allow the observation of the entanglement phenomenon, and the measurement of quantum magic. 
From the measured spin correlation at the \ttbar~production threshold and high \ttbar~mass, entanglement is observed with a large fraction of the \ttbar~decays being spacelike separated. The observation of entanglement in \ttbar~events with two high transverse momentum leptons of opposite charge is also presented. Finally, the first TeV-scale experimental measurement of quantum magic, an important variable for the characterization of quantum states in quantum information science, is presented. These measurements provide one of the first connections between quantum information science and particle physics, and show the potential of collider experiments in the studies of the foundations of quantum mechanics.

}

\FullConference{The European Physical Society Conference on High Energy Physics (EPS-HEP2025)\\
7-11 July 2025\\
Marseille, France\\}

\begin{document}

\renewcommand{\hookAfterAbstract}{%
\par\bigskip
}

\renewcommand{\logo}{\relax}

\maketitle


\section{Introduction}
The top quark ($t$) has the largest mass ($m_t$) among the elementary particles known to date~\cite{ParticleDataGroup:2024cfk}. It has a very short lifetime, $\approx10^{-25}~\text{s}$, shorter than the quantum chromodynamics (QCD) hadronization time scale ($1/\Lambda_{QCD}\approx10^{-24}~\text{s}$), and the spin de-correlation time scale ($m_t/\Lambda^2_{QCD}\approx10^{-21}~\text{s}$), in the top quark -- antiquark ($\bar{t}$) pair ($\ttbar$) system~\cite{Bigi:1986jk,Mahlon:2010gw}. Therefore, $t$ and $\bar{t}$ spins stay correlated and top quark polarization and spin-correlations in \ttbar~production could be measured from the angular distributions of the decay products. The first evidence for \ttbar~spin correlation was published by the D0 experiment at the Tevatron~\cite{D0:2011kcb,D0:2015kta}, and the ATLAS and CMS experiments at the LHC have made several top quark polarization and \ttbar~spin correlation measurements using dilepton~\cite{ATLAS:2012ao,CMS:2013roq,ATLAS:2014abv,ATLAS:2014aus,CMS:2016piu,CMS:2019nrx,ATLAS:2019zrq,ATLAS:2023fsd} and lepton+jets~\cite{ATLAS:2014aus,CMS:2015cal} final states. These measurements provide stringent tests of the standard model (SM). In specific phase space regions, they also allow novel tests of quantum theory, such as entanglement~\cite{Afik:2020onf}, Bell inequalities~\cite{Fabbrichesi:2021npl}, and magic~\cite{White:2024nuc}, at the high energy regime using unstable particles that are not possible in low-energy experiments using electrons or photons. 
This proceedings presents the latest \ttbar~spin-correlation results~\cite{CMS:2024zkc} together with the corresponding quantum-entanglement~\cite{CMS:2024zkc,CMS:2024pts} and quantum-magic~\cite{CMS:2025cim} measurements performed with the CMS detector~\cite{CMS:2008xjf}.

\section{Full spin density matrix measurement with lepton+jets events}
\label{sec:lepton_plus_jets}
Measurements of the polarization and spin correlation of \ttbar~in the lepton+jets channels are performed by the CMS experiment~\cite{CMS:2024zkc}, using the helicity basis~\cite{Bernreuther:2015yna}.  
The \ttbar~differential cross section can be written as
\begin{align}
\label{eq:diff_cs}
\sum_{tot}\left(\phi_{P(\bar{P})}),\theta_{P(\bar{P})} \right) &=\frac{d^4\sigma}{d\phi_Pd\cos(\theta_P)d\phi_{\bar{P}}d\cos(\theta_{\bar{P}})} \\\nonumber
 & = \sigma_{norm}\left(1+\kappa \vect{P}\cdot\vect{\Omega}+\bar{\kappa}\vect{\bar{P}}\cdot\vect{\bar{\Omega}}-\kappa\bar{\kappa}\vect{\Omega}\cdot(\mathbf{C}\vect{\bar{\Omega}})\right)\\\nonumber
 & = \Sigma_0+\sum_{m=1}^{15}Q_m\Sigma_m
\end{align}
where, $\vect{P}$ is the polarization vector, $\mathbf{C}$ the $3\times3$ spin-correlation matrix,  
$\kappa$ the spin-analyzing power, and $\sigma_{norm}$ is the overall normalization. 
The unit vector, $\vect{\Omega}$, in the helicity basis
is defined as $\Omega(\bar{\Omega})=(\sin(\theta_{p(\bar{p})})\cos(\phi_{p(\bar{p})}),\sin(\theta_{p(\bar{p})})\sin(\phi_{p(\bar{p})}),\cos(\theta_{p(\bar{p})}))$
where $\phi_{p(\bar{p})}$ is the azimuthal and $\theta_{p(\bar{p})}$ the polar angle of the decay product $p(\bar{p})$ of the top (anti)quark. 
The spin dependence of \ttbar~is fully characterized by the 15 coefficients $Q_m=\{P_n,P_r,P_k,\bar{P_n},...,C_{nn},C_{nr},...,C_{kk}\}$ which are all measured by the angular distributions 
\begin{equation}
\Sigma_m=\sigma_{norm}\{\kappa\sin\theta_P\cos\phi_P,...,\kappa\bar{\kappa}\cos\theta_P\cos\theta_{\bar{P}}\}.
\end{equation}
In the lepton+jets channel, events with a lepton and down-type quark (from the W boson) yield the maximum top quark spin transfer to the decay products, with $\kappa\to1$. 
The \ttbar~signal is generated using \Powheg v2~\cite{Nason:2004rx,Frixione:2007vw,Frixione:2007nw} 
interfaced with \PYTHIA~\cite{Sjostrand:2014zea} with electroweak corrections included calculated from \Hathor~\cite{Aliev:2010zk} and uncertainties for higher order QCD effects are estimated using \Powheg-MiNNLO~\cite{Mazzitelli:2020jio}. 
For all simulated samples, the CP5 underlying event tune~\cite{CMS:2019csb} is used.

A neural network (NN) is used to reconstruct the \ttbar~system and to provide down-type quark identification. 
The inputs to the NN are variables related to lepton kinematic distributions, missing energy, jet kinematics, and b-tagging scores. 
For each event, every possible assignment of detector-level jets to the \ttbar~decay products is passed to the NN, and the permutation that produces the highest NN score (S$_\text{NN}$) is used. 
Event categories used in the extraction of the signal are based on lepton flavor, number of b-tags, and S$_\text{NN}$ score.  
Events with low NN scores, S$_\text{NN}<0.1$, are discarded since they contain only a low fraction of correctly reconstructed \ttbar~events.  
High S$_\text{NN}$ score is defined separately for events with exactly one (1b) or both (2b) jets identified as b jets from the \ttbar~decay   as  S$_\text{high}(1b)>0.30$ and S$_\text{high}(2b)>0.36$, respectively. These thresholds are optimized to minimize the uncertainties in the spin density matrix. 
The fraction of correctly assigned jets, including down-type quark identification, is $\sim$40--50\% for S$_\text{high}(2b)$. 

\begin{figure}
    \centering
    \includegraphics[width=14.0cm]{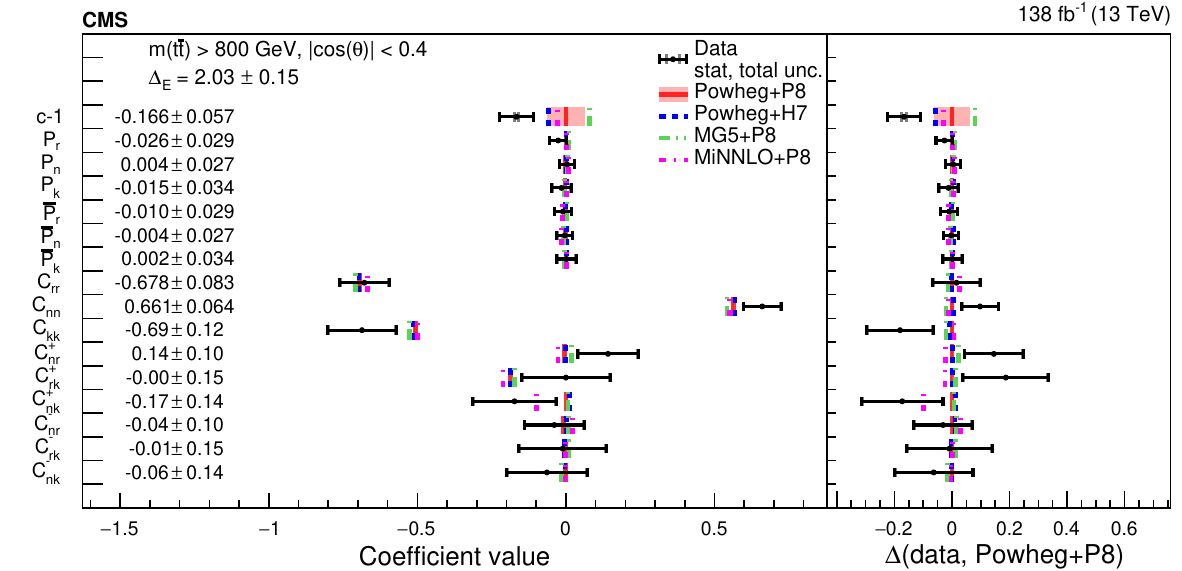}
    \caption{Measured coefficients of the full matrix for m(\ttbar) $>800~\text{GeV}$ and $\lvert\cos\theta\rvert<0.4$. The data are shown with the statistical uncertainty (inner error bars) and total uncertainty (outer error bars) and compared to the predictions. In the right panel, results are presented with the \Powheg+\PYTHIA~predictions subtracted. The \Powheg+\PYTHIA~prediction is displayed with QCD scale and PDF uncertainties. Figure taken from Ref.~\cite{CMS:2024zkc}.}
    \label{fig:SC_matrix}
\end{figure}

The measurement of the coefficients $Q_m$ are made using detector-level templates that are fit to the data. Each template describes solely the effect of $Q_m$. 
Spin density matrix coefficients are measured inclusively and differentially in bins of the invariant mass of the \ttbar~system (m(\ttbar)), top quark scattering angle 
$\lvert\cos\theta\rvert$, and the transverse momentum of the top quarks, p$_\text{T}$(t). 
The coefficients are extracted simultaneously using a binned likelihood fit in various regions of phase space to reconstructed-level templates simultaneously fitting the data in 16 categories; $(2b,1b)\times(\text{S}_\text{high},\text{S}_\text{low})\times$ 4 data-taking periods. 
The measured coefficients for m(\ttbar) $>800~\text{GeV}$ and $\lvert\cos\theta\rvert<0.4$ are displayed in Fig.~\ref{fig:SC_matrix}.
Good agreement is observed between the central values of the data and the ones predicted by the SM, and also  the measurements by CMS in the dilepton channel~\cite{CMS:2019nrx}.

\section{Observation of quantum entanglement from \ttbar~ spin density matrix}
\label{sec:entanglement_ljets}
As any spin-1/2 particle, the quantum state of the top quark can be described as a {\it qubit}, i.e. a two-state quantum system. Therefore, \ttbar~is a two-qubit system, and because the $t$ and $\bar{t}$ spins are correlated, a \ttbar~pair represents an entangled quantum state.  
The ATLAS and CMS experiments observed \ttbar-entanglement at the production threshold in the dilepton channel~\cite{ATLAS:2023fsd,CMS:2024pts}. 
CMS experiment also observed entanglement (assuming no local hidden variable theory) in the lepton+jets channel, extending the measurement beyond the production-threshold region to include events with high m(\ttbar)~\cite{CMS:2024zkc}. 
The Peres-Horodecki criterion~\cite{Peres:1996dw, Horodecki:1997vt} provides the sufficient condition for a quantum state to be separable. From this, a sufficient condition for entanglement using the diagonal elements of the spin correlation matrix is obtained~\cite{Afik:2020onf,Afik:2022kwm} as $\Delta_E=C_{nn}+|C_{rr}+C_{kk}|>1$. At \ttbar~production threshold, $\text{m}(\ttbar)\sim2\text{m}(t)$. In this region, the differential cross section is given by
\begin{equation}
\frac{d\sigma}{d\cos\chi}=\sigma_{norm}(1+D\kappa\bar{\kappa}\cos\chi)
\end{equation}
where $D=-Tr[C]/3$. In this region of phase space, for $gg$ fusion events,  $C_{rr},C_{kk}>0$~\cite{Baumgart:2012ay} and therefore the entanglement criterion becomes
    $\Delta_E=-3D=Tr[C]>1$.
The value of $D$ is extracted using the opening angle  between two decay products $\cos\chi=\Omega\cdot\bar{\Omega}$ in the helicity basis, which is sensitive to entanglement in the spin-singlet state expected from $gg\to\ttbar$~production.
At high m(\ttbar), and low $\lvert\cos\theta\rvert$, for both in $q\bar{q}$ annihilation and $gg$ fusion events, entanglement in a spin-triplet state could be investigated with $\tilde{D}=\frac{1}{3}(C_{nn}-C_{rr}-C_{kk})$~\cite{Baumgart:2012ay,Aguilar-Saavedra:2022uye}.
When $\text{p}_\text{T}(t)\sim\text{m}_t$, $C_{rr}$ and $C_{kk}$ are negative~\cite{Baumgart:2012ay} and the condition for entanglement reduces to $\Delta_E=3\tilde{D}>1$.
The quantity $\tilde{D}$ is extracted using $\tilde{\chi}=-\Omega_n\bar{\Omega}_n+\Omega_r\bar{\Omega}_r+\Omega_k\bar{\Omega}_k$.

\begin{figure}
    \centering
    \includegraphics[width=6.9cm]{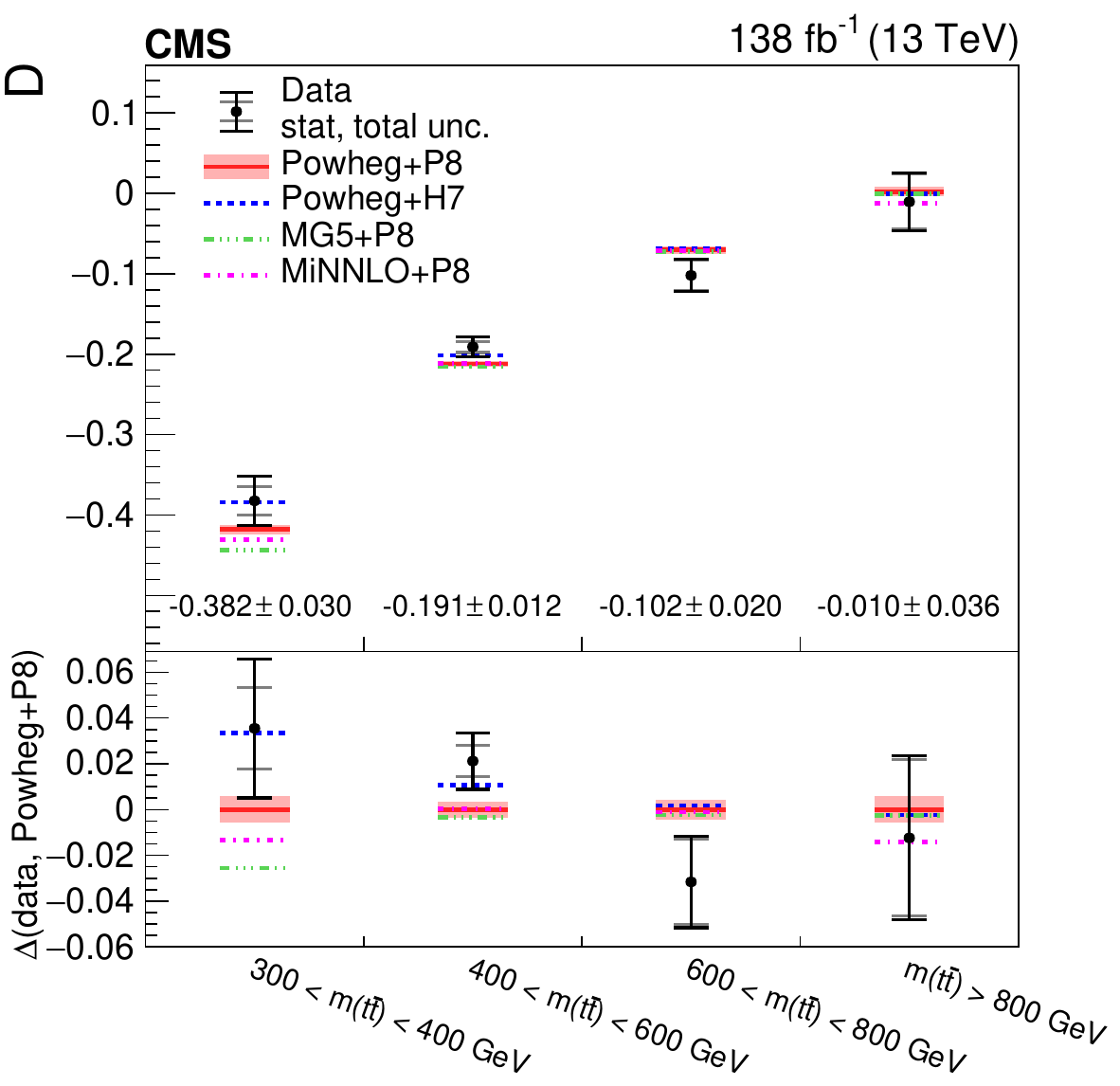}
    \includegraphics[width=6.9cm]{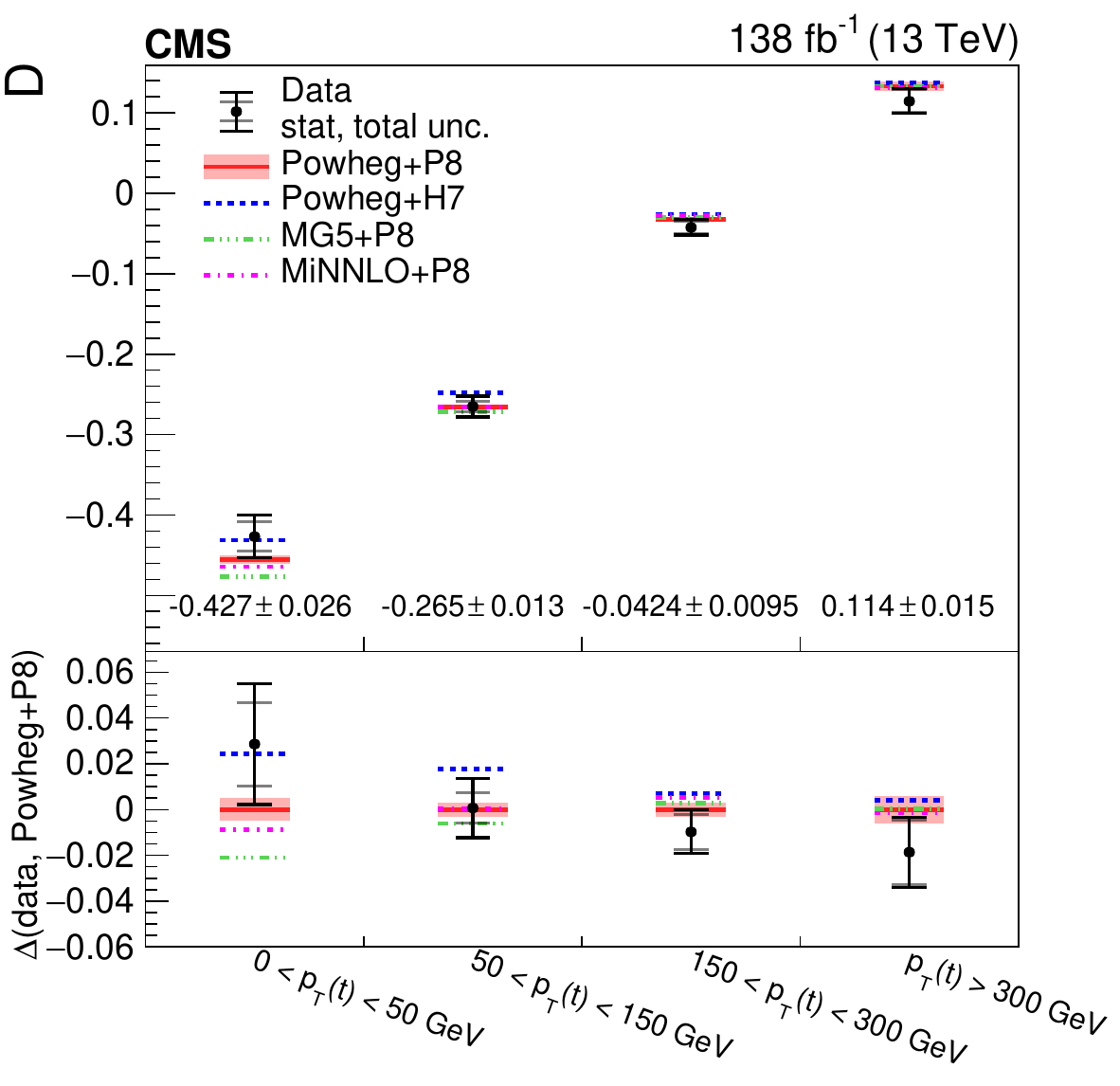}
    \caption{Measurements of $D$ in bins of m(\ttbar) and p$_\text{T}$(t) shown with statistical (inner error bars) and total uncertainty (outer error bars) are compared to various predictions. In the lower panel, the results are presented with the \Powheg+\PYTHIA~ predictions subtracted. The \Powheg+\PYTHIA~ predictions are displayed with QCD scale and PDF uncertainties. Figures taken from Ref.~\cite{CMS:2024zkc}.}
    \label{fig:D_ljets}
\end{figure}

\begin{figure}
    \centering
    \includegraphics[width=8.0cm]{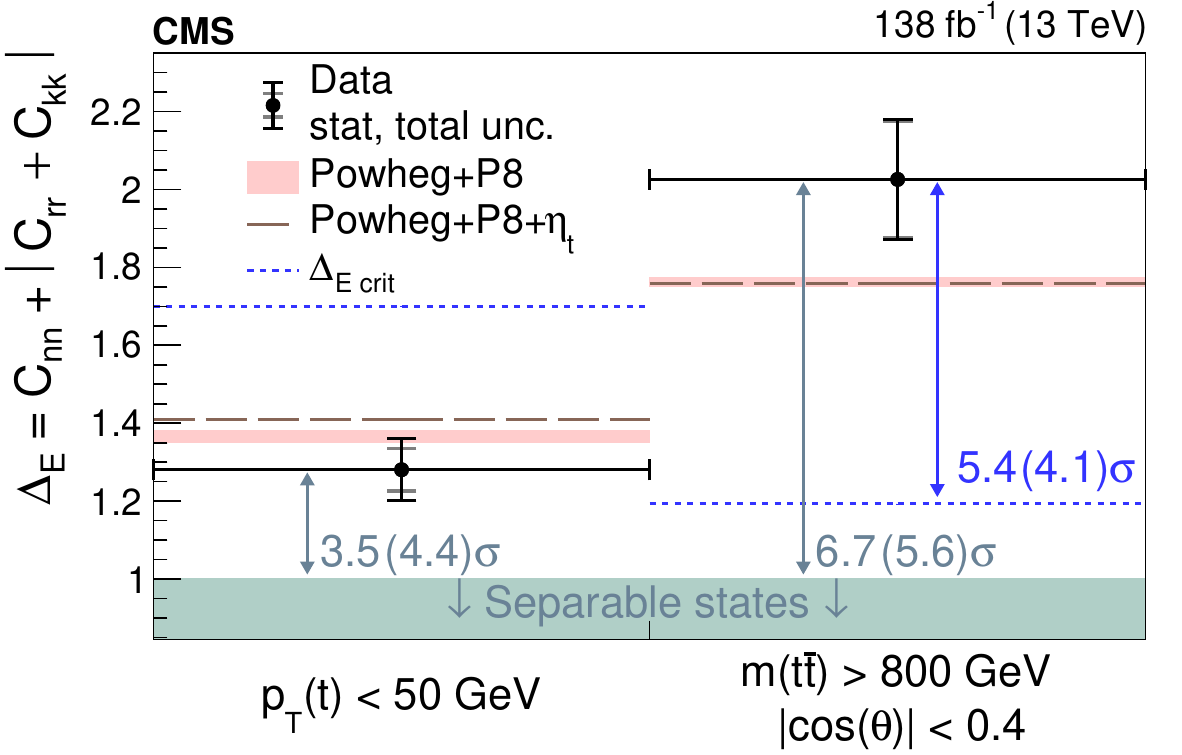}
    \caption{The observed levels of entanglement ($\Delta_E$) are displayed in the \ttbar~threshold region using the $D$ measurement (first bin), and in the high-m(\ttbar) region using the full matrix measurement (second bin). The data are shown with the statistical uncertainty (inner error bars) and total uncertainty (outer error bars) and compared to the predictions including and excluding the $\eta_t$ state. The prediction without $\eta_t$ is displayed with the QCD scale and PDF uncertainties. The horizontal dashed blue lines indicate the maximum level of entanglement $\Delta_{E,crit}$ that can be explained by the exchange of information between $t$ and $\bar{t}$ at the speed of light. Figure taken from Ref.~\cite{CMS:2024zkc}.}
    \label{fig:entanglement_ljets}
\end{figure}

The measurements of $D$ and $\tilde{D}$, displayed in Fig.~\ref{fig:D_ljets} are made in bins of m(\ttbar) and p$_\text{T}$(t). In all cases, good agreement with predictions are observed. The observed entanglement strengths are displayed in Fig.~\ref{fig:entanglement_ljets}: for $\text{p}_\text{T}(t)<50~\text{GeV}$ using the $D$ measurement, and for m(\ttbar) $>800~\text{GeV}$ with $\lvert\cos\theta\rvert<0.4$ using the full matrix measurement. The former, yielding an observed (expected) significance of 3.5 (4.4) standard deviations ($\sigma$), is less sensitive than that of the dilepton channel analysis~\cite{CMS:2024pts}. 
In the latter, the observed (expected) significance is 6.7 $\sigma$ (5.6 $\sigma$). 
In this bin the significance is also calculated with respect to the lower boundary of $\Delta_{\text{E}~\text{crit}}$~\cite{Demina:2024dst}, which is the maximum level of entanglement that could be explained classically by the exchange of information between $t$ and $\bar{t}$ at the speed of light.
The observed (expected) significance with respect to the lower boundary of $\Delta_{\text{E}~\text{crit}}$,
corresponding to a spacelike-separated fraction of events of 
90\%~\cite{Severi:2021cnj}, is 5.4 $\sigma$ (4.1 $\sigma$). 
This represents the first observation of quantum entanglement at high m(\ttbar).

\section{Observation of quantum magic states of top quark pairs}
Quantum magic quantifies the computational advantage of quantum states over classical states or a quantum computer over a classical computer. For mixed states like the two-qubit \ttbar~ system, in the helicity basis $\{n,r,k\}$, the magic, based on R\'enyi entropy~\cite{Leone:2021rzd}, is given by Ref.~\cite{White:2024nuc} as
\begin{equation} 
\tilde{M}_2=-\log_2\left(\frac{1+\Sigma_{i\in n,k,r}[(P_i^4+\bar{P}_i^4)]+\Sigma_{i,j\in n,k,r}C_{ij}^4}{1+\Sigma_{i\in n,k,r}[(P_i^2+\bar{P}_i^2)]+\Sigma_{i,j\in n,k,r}C_{ij}^2} \right)  
\end{equation}
with $P_i(\bar{P}_i)$ representing the coefficients of the  polarization vector $\vect{P}$, and $C_{ij}$ the spin correlation coefficients, as in Eqn.~\ref{eq:diff_cs}.  
Magic is a property of quantum states and zero magic corresponds to a classical computer. $\tilde{M}_2$ is nonlinear, phase-space dependent, and therefore, the magic observation in the process $\text{pp}\to\ttbar$~does not imply the existence of quantum magic in the individual partonic processes $q\bar{q}\to\ttbar$~and $gg\to\ttbar$.

\begin{figure}
    \centering
    \includegraphics[width=6.2cm]{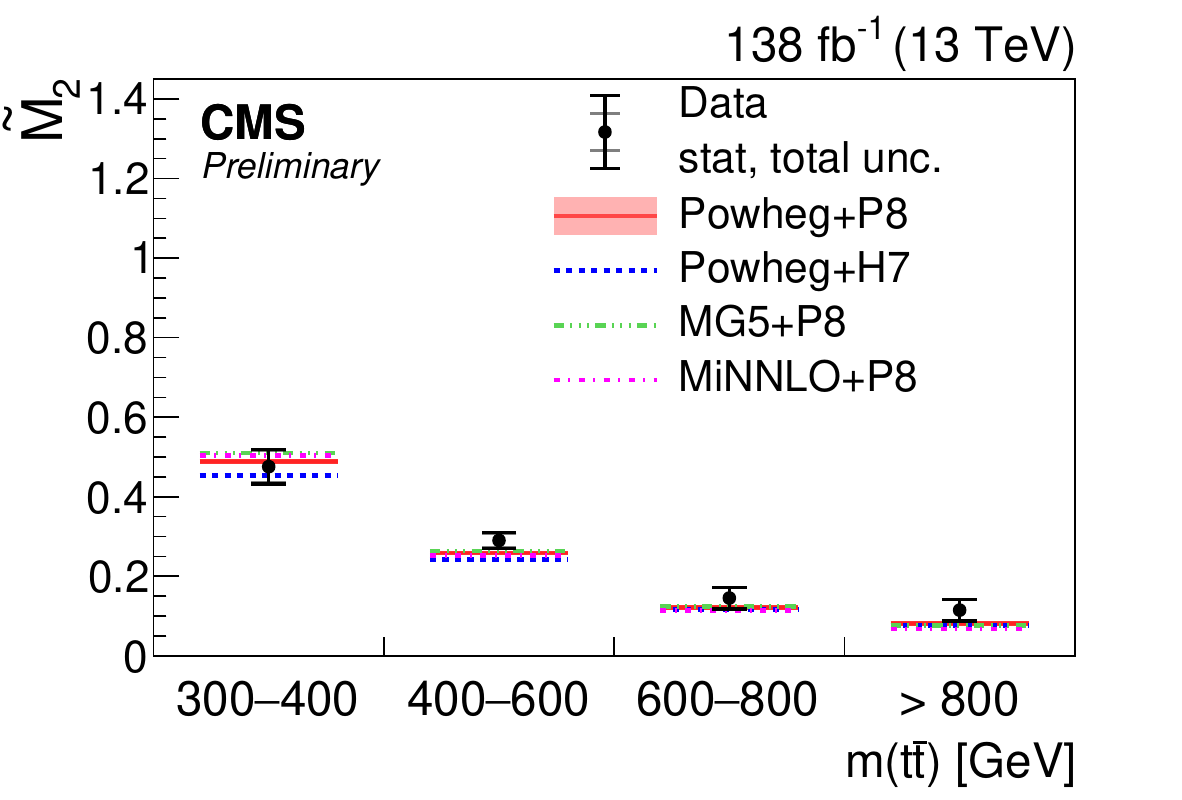}
    \includegraphics[width=6.2cm]{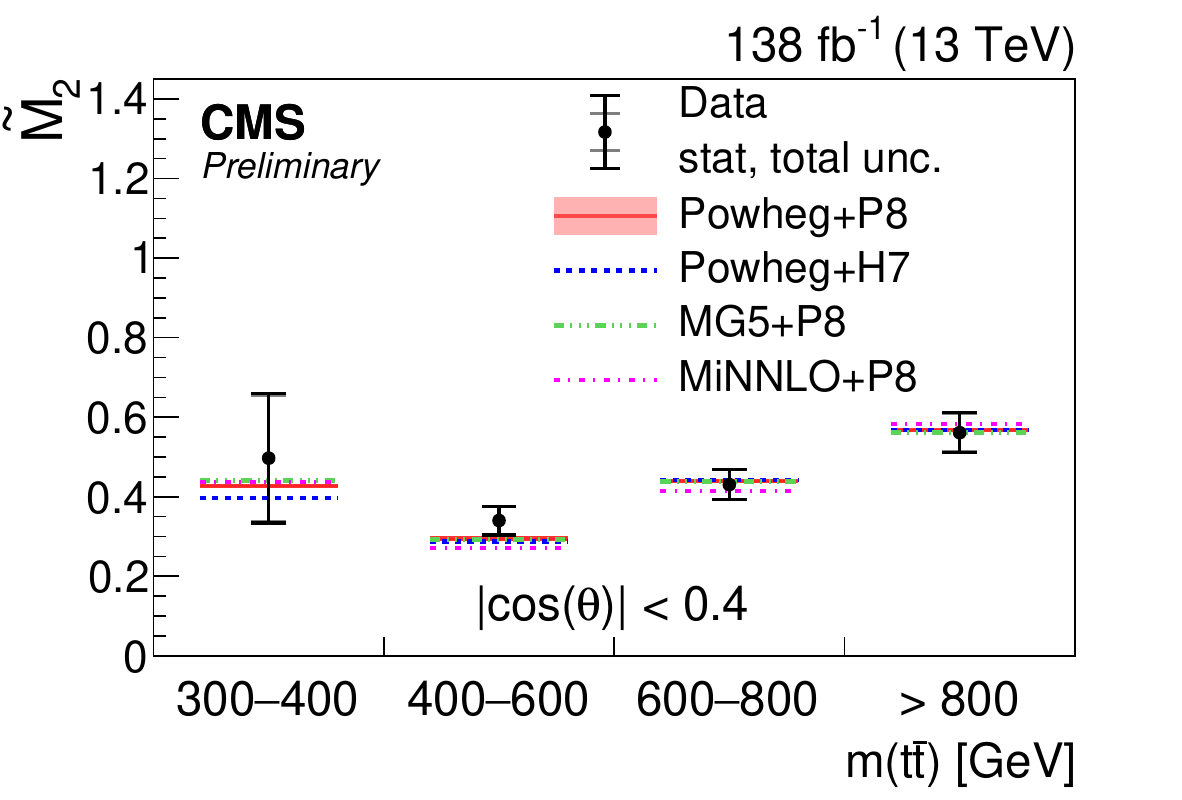}
    \includegraphics[width=6.2cm]{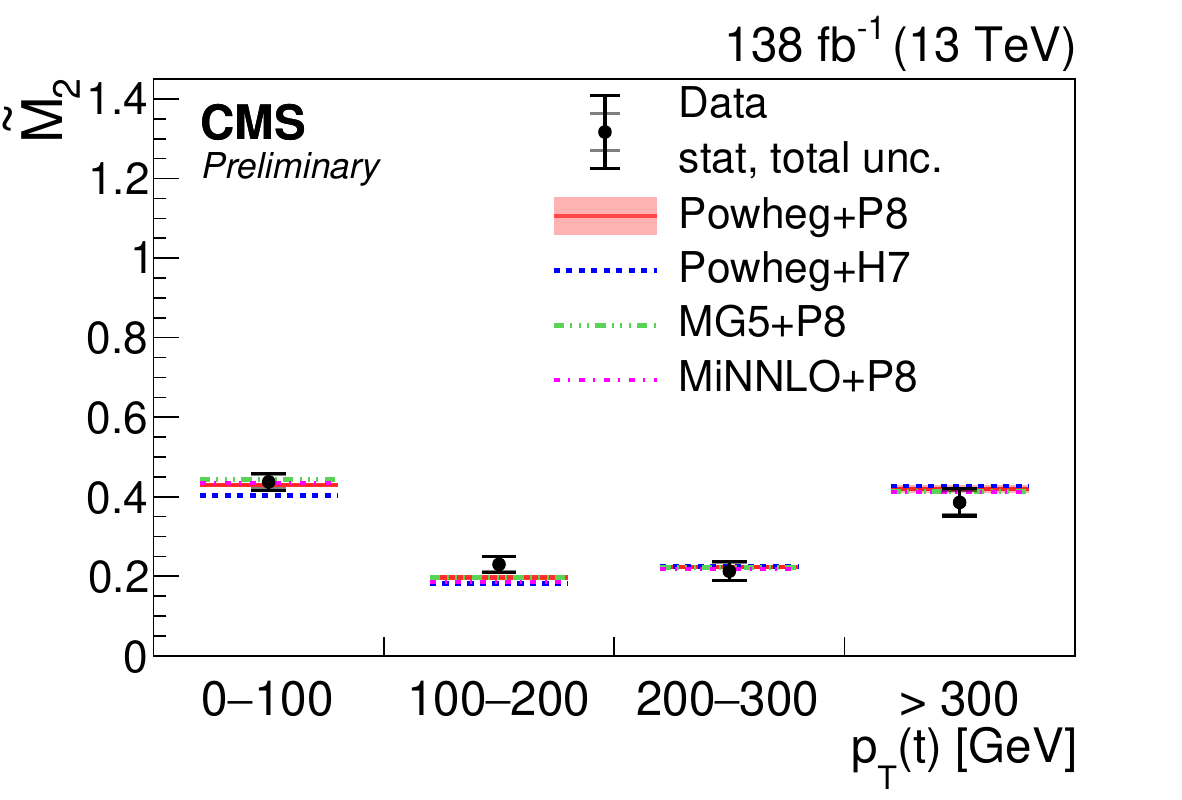}
    \caption{Results of $\tilde{M}_2$ measurements compared to the predictions from various simulation configurations. The measurements are made in bins of m(\ttbar) (upper left),  m(\ttbar) with $\lvert\cos\theta\rvert<0.4$ (upper right), and p$_\text{T}$(t) (lower). The \Powheg+\PYTHIA~ predictions are displayed with the QCD scale and parton distribution function uncertainties, while for all other predictions only the central values are displayed. Figures taken from Ref.~\cite{CMS:2025cim}.}
    \label{fig:magic}
\end{figure}

The CMS analysis~\cite{CMS:2025cim} measures $\tilde{M}_2$ of \ttbar~pairs in different phase space regions using the spin density matrix measurement in the \ttbar~ lepton+jets channel~\cite{CMS:2024zkc} presented in Sec.~\ref{sec:lepton_plus_jets}. 
Fig.~\ref{fig:magic} displays the measured $\tilde{M}_2$ in bins of m(\ttbar) (also with $\lvert\cos\theta\rvert<0.4$), and of p$_\text{T}$(t). The measurements are  compared to the predictions of the SM using various simulation setups.  The results depend on the phase space region; $\tilde{M}_2$ decreases monotonically from m(\ttbar) = 300--400 GeV to  m(\ttbar) $>800$ GeV but when the condition $\lvert\cos\theta\rvert<0.4$ is imposed it remains approximately constant between first and the last bins considered. The measurement in bins of p$_\text{T}$(t) shows a decrease from 0--100 GeV to 100--300 GeV, and then an increase  for $>300~\text{GeV}$. In all bins, a good agreement is observed between the data and the predictions. In the measurements, statistical uncertainties are dominant.
This represents the first positive quantum magic measurement from \ttbar~spin density matrix. 

\section{Observation of quantum entanglement in the dilepton channel}
\begin{figure}
    \centering
    \includegraphics[width=8.2cm]{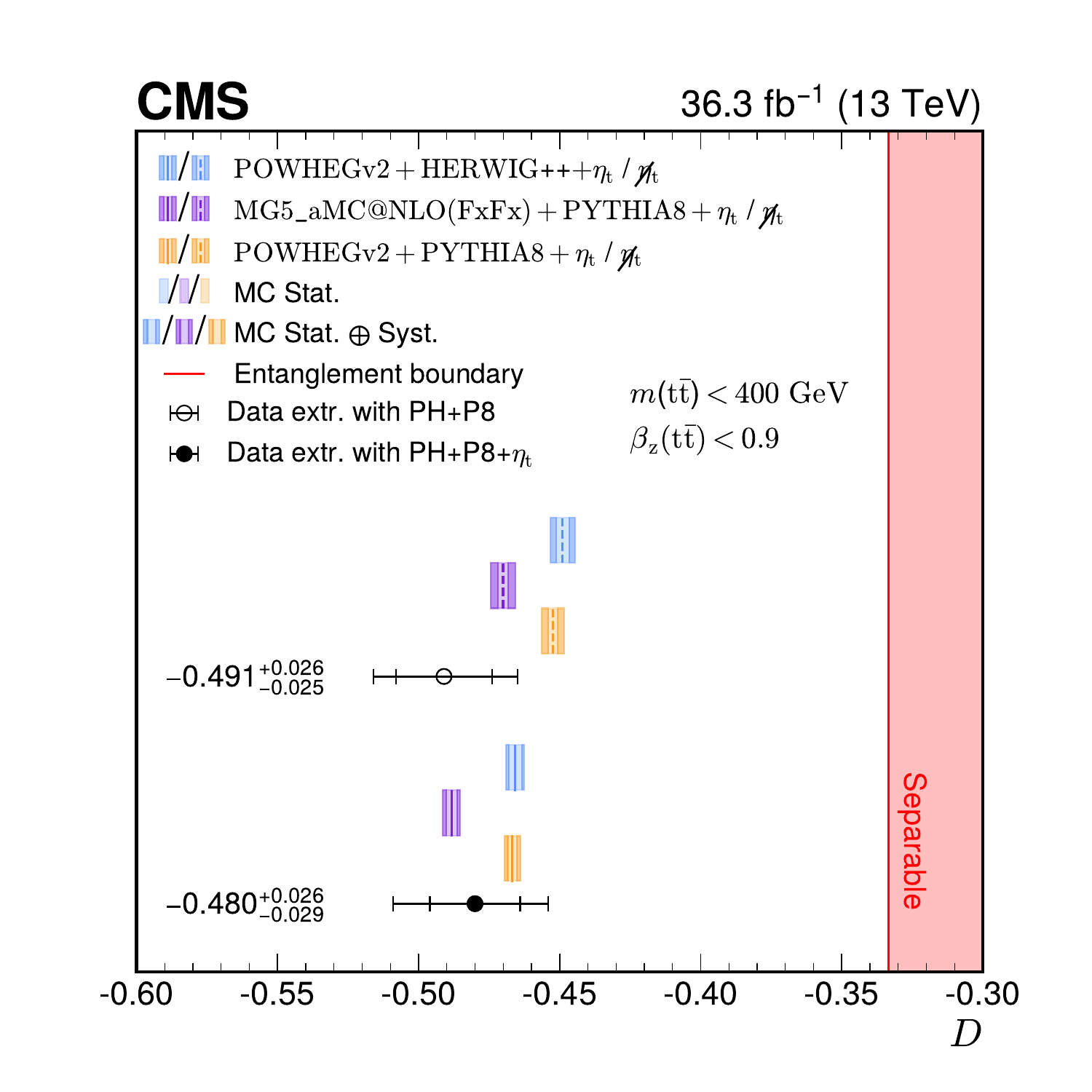}
    \caption{Measurements of $D$ (black filled or open point) compared with the predictions from simulations including (solid line) or not including (dashed line) the $\eta_t$ state contributions. The predictions without the $\eta_t$ state is shown with $\cancel\eta_t$. Inner (outer) error bars represent the statistical (total) uncertainty for data. The statistical uncertainty in the predictions is shown by the light shaded region and the total uncertainty, including scale and parton distribution function uncertainties, is represented by the darker shaded region. The boundary for entanglement is indicated by the shaded region at high values of $D$. Figure taken from Ref.~\cite{CMS:2024pts}.}
    \label{fig:dilepton_entanglement_D}
\end{figure}

The level of entanglement in \ttbar~events is measured by CMS at 13 TeV with the data corresponding to an integrated luminosity of 36.3 fb$^{-1}$ using events with two high $\text{p}_\text{T}$ leptons of opposite sign~\cite{CMS:2024pts}. 
The \ttbar~differential cross section can be written as  
\begin{equation}
    \frac{d\sigma}{d\cos\varphi}=\sigma_{norm}(1-D\cos\varphi)
\end{equation}
with $\cos\varphi=\hat{\ell}^+\cdot\hat{\ell}^-$ defined by  the angle between the two charged leptons in their respective parent top quark rest frames. The parameter $D$ is an entanglement-sensitive observable with $D=0$ correspond to the case with no \ttbar~spin correlation. The parameter $D$ is related to the diagonal coefficients of the spin density matrix $\mathbf{C}$ as $D=Tr[C]/3$, which combined with the Peres-Horodecki criterion (see Sec.~\ref{sec:entanglement_ljets}) yields the condition for an entangled \ttbar~system: $D<-1.3$. 
The parameter $D$ is measured using the two--bin asymmetry variable with $D=-2A_D$, where
\begin{equation}
    A_D=\frac{N(\cos\varphi>0)-N(\cos\varphi<0)}{N(\cos\varphi>0)+N(\cos\varphi<0)}.
\end{equation}
The measurement of the $D$ is made using a binned-profile likelihood fit to $\cos\varphi$ in the most sensitive phase space, i.e. the relative velocity between the laboratory and \ttbar~frames, $\beta_z(\ttbar)=|p_z^t+p_z^{\bar{t}}|/|E^t+E^{\bar{t}}|<0.9$ and $345<$m(\ttbar)$<400$ GeV~\cite{CMS:2024pts}.
Variations of $D$ are obtained using samples with different degrees of SM and no-spin correlation assumptions. 
Separate fits are made including and excluding a ground state toponium ($\eta_t$) that is recently observed by the CMS experiment~\cite{CMS:2025kzt} and confirmed by the ATLAS experiment~\cite{ATLAS:2025mvr}. The $\eta_t$ state is modeled as a non-relativistic QCD quasi-bound state with an invariant mass of 343 GeV and a cross section of 6.4 pb.
The results are shown in Fig.~\ref{fig:dilepton_entanglement_D}. 
The fit yields an observed value of $D=-0.480^{+0.026}_{-0.029}$ with an expected value of $D=-0.467^{+0.026}_{-0.029}$ including the predicted $\eta_t$ state. This corresponds to an observed (expected) significance of the top quark entanglement observation of 5.1$\sigma$ (4.7$\sigma$). The fit without the  $\eta_t$ state yields $D=-0.491^{+0.026}_{-0.025}$ with an expected value of $D=-0.452^{+0.025}_{-0.026}$ corresponding to an observed (expected) significance of 6.3 $\sigma$ (4.7 $\sigma$). Therefore, irrespective of the existence of $\eta_t$, the entanglement in \ttbar~ is observed with $>5~\sigma$ significance, and including the $\eta_t$ state in the simulation further improves  the agreement between the data and the simulation.  

\clearpage
\newpage

\bibliography{main.bib}
\end{document}